\documentclass[fleqn,usenatbib]{mnras}

\usepackage{newtxtext,newtxmath}
\usepackage[T1]{fontenc}
\usepackage{ae,aecompl}

\usepackage{graphicx}	% Including figure files
\usepackage{amsmath}	% Advanced maths commands

\usepackage{amssymb}	% Extra maths symbols
\usepackage{gensymb}
\usepackage{threeparttable} 
\usepackage{subfigure}
\usepackage{enumerate}

\usepackage{lipsum,pdflscape}
\usepackage{url}
\usepackage{breakurl}
\usepackage{hyperref}

\newcommand{\HI}{{\sc H\,i }}
\title[Identifying thin discs in MaNGA galaxies]{An analytical model to kinematically identify thin discs in MaNGA galaxies}

\author[M. Yang et al.]{Meng Yang$^{1,2}$\thanks{E-mail: myang@shao.ac.cn}, 
Anne-Marie Weijmans$^{1}$,
Matthew A. Bershady$^{3,4,5}$,\newauthor 
Michael Merrifield$^{6}$,
Nicholas F. Boardman$^{7}$,
Niv Drory$^{8}$
\\
$^{1}$School of Physics and Astronomy, University of St Andrews, North Haugh, St Andrews, KY16 9SS, UK\\
$^{2}$Shanghai Astronomical Observatory, Chinese Academy of Sciences, 80 Nandan Road, Shanghai 200030, China\\
$^{3}$Department of Astronomy, University of Wisconsin-Madison, Madison, WI 53706, USA\\
$^{4}$South African Astronomical Observatory, PO Box 9, Observatory 7935, Cape Town, South Africa\\
$^{5}$Department of Astronomy, University of Cape Town, Private Bag X3, Rondebosch 7701, South Africa \\
$^{6}$School of Physics and Astronomy, University of Nottingham, University Park, Nottingham, NG7 2RD, UK\\
$^{7}$Department of Physics and Astronomy, University of Utah, Salt Lake City, UT 84112, USA\\
$^{8}$McDonald Observatory, The University of Texas at Austin, 1 University Station, Austin, TX 78712, USA
}

\date{Accepted XXX. Received YYY; in original form ZZZ}

\pubyear{2020}

\begin{document}
\label{firstpage}
\pagerange{\pageref{firstpage}--\pageref{lastpage}}
\maketitle

% Abstract of the paper
\begin{abstract}
We present an analytical model to identify thin discs in galaxies, and apply this model to a sample of SDSS MaNGA galaxies. This model fits the velocity and velocity dispersion fields of galaxies with regular kinematics. By introducing two parameters $\zeta$ related to the comparison of the model's asymmetric drift correction to the observed gas kinematics and $\eta$ related to the dominant component of a galaxy, we classify the galaxies in the sample as "disc-dominated", "non-disc-dominated", or "disc-free" indicating galaxies with a dominating thin disc, a non-dominating thin disc, or no thin disc detection with our method, respectively. The dynamical mass resulting from our model correlates with stellar mass, and we investigate discrepancies by including gas mass and variation of the initial mass function. As expected, most spiral galaxies in the sample are disc-dominated, while ellipticals are predominantly disc-free. Lenticular galaxies show a dichotomy in their kinematic classification, which is related to their different star formation rates and gas fractions. We propose two possible scenarios to explain these results. In the first scenario, disc-free lenticulars formed in more violent processes than disc-dominated ones, while in the second scenario, the quenching processes in lenticulars lead to a change in their kinematic structures as disc-dominated lenticulars evolve to disc-free ones. 
\end{abstract}

% Select between one and six entries from the list of approved keywords.
% Don't make up new ones.
\begin{keywords}
galaxies: kinematics and dynamics -- galaxies: structure -- galaxies: disc -- galaxies: evolution
\end{keywords}

%%%%%%%%%%%%%%%%%%%%%%%%%%%%%%%%%%%%%%%%%%%%%%%%%%

%%%%%%%%%%%%%%%%% BODY OF PAPER %%%%%%%%%%%%%%%%%%

\section{Introduction}

Galaxy morphology classification is a fundamental tool for research into galaxy formation and evolution. Since Hubble first arranged galaxy images into a tuning fork~\citep{1926ApJ....64..321H}, several further classification schemes have been developed~\citep[e.g.][]{1959HDP....53..275D,1958PASP...70..364M}. In these schemes, galaxies are mostly classified by their components: spiral galaxies contain an extended exponential disc with clear spiral arm structure and usually a concentrated centre regarded as a bulge, while elliptical galaxies are mostly structurally smooth without distinguishable features. Lenticular or S0 galaxies share the smooth structure of ellipticals, but do have a clear disc structure. 

Generally, visual classification is subject to observational bias and is a time-consuming process. Moreover, the rise of large galaxy surveys makes it almost impossible to classify galaxy morphologies by eye. A series of techniques have therefore been developed to automate this process. Photometric parameters, such as S\'ersic index~\citep{1968adga.book.....S}, are widely used for quantitative morphology classification. \citet{2003ApJS..147....1C} developed the CAS classification systems which has as quantitative parameters the concentration (C), asymmetry (A), and clumpiness (S) of a galaxy. Making use of large numbers of amateur volunteers instead of professionals to classify by eye has also proved to be an efficient and effective way to obtain galaxy classifications \citep[e.g. Galaxy Zoo:][]{2008MNRAS.389.1179L,2013MNRAS.435.2835W}. Recently, the development of deep learning algorithms has also begun to contribute to automated galaxy morphology classification~\citep[e.g.][]{2015ApJS..221....8H,2018MNRAS.476.3661D}. 

Galaxy images alone, however, do not provide the intrinsic information of individual galaxy components, and therefore should be complemented by kinematic information to build a complete picture of a galaxy's structure. 
One straightforward way to utilise galaxy kinematic information is through well-defined parameters, such as $\lambda_{Re}$, which is defined as the quantified projected stellar angular momentum per unit mass. Introduced by~\citet{2004MNRAS.352..721E}, this parameter is used to study the intrinsic structure of galaxies, and separate early-type galaxies into slow and fast rotators, which are physically distinguished by their dominant motions of stars. This parameterisation has been applied to classify large samples of galaxies~\citep[e.g.][]{2019A&A...632A..59F,2017ApJ...835..104V}.

Dynamical modelling is another way to study kinematic properties and to recover individual galaxy components. The collisionless Boltzmann equation allows us to describe the steady state of stellar systems with distribution functions (DF)~\citep{1915MNRAS..76...70J}. The stars in different galaxy components populate different orbital states, which means they can be distinguished by dynamical modelling. Although numerical dynamical modelling methods are powerful for decomposing galaxy kinematic components~\citep[e.g][]{2018MNRAS.479..945Z}, these methods can be quite computationally intensive.

An alternative way to study galaxy kinematic properties is through analytical models, which is less time consuming compared to numerical dynamical modelling, and can provide additional information on kinematic parameters.
A number of analytical models were developed to describe DFs for different systems, such as the Osipkov-Merritt Models~\citep{1979PAZh....5...77O,1985AJ.....90.1027M} for anisotropic spherical systems and the Evans model~\citep{1994MNRAS.271..202E} for axisymmetric systems. 
For disc galaxies with intermediate inclinations, a simple analytical model is built to extract the shape of stellar velocity ellipsoid from line-of-sight velocity dispersions~\citep{1997MNRAS.288..618G,2000MNRAS.317..545G,2008MNRAS.388.1381N}.
The asymmetric drift phenomenon, which is the difference between the mean stellar tangential velocity and the circular velocity, caused by the stellar density and velocity dispersion gradient~\citep{2008gady.book.....B}, also offers us a scheme to help build analytical models. Quantifying asymmetric drift in an axisymmetric system provides a possible way to identify the dynamical states of galaxies ~\citep[e.g.][]{2010ApJ...716..234B}.

In this paper, we build an analytical model based on the asymmetric drift correction in the Evans model to identify thin discs, which are well described by the thin-disc approximation~\citep{2008MNRAS.383.1343W}, in galaxies. We classify a sample of galaxies with regular rotating features as disc-dominated, disc-free and non-disc-dominated according to whether the model holds, and then study their kinematic properties and morphologies. The paper is organised in the following way: we introduce the observations and sample selection in Section~\ref{sec:obs}, and in Section~\ref{sec:methods} we describe our model in detail, and introduce a morphological classification based on our model. We show the results of applying our model to the sample in Section~\ref{sec:modelfitness}. The galaxy properties of these different morphology classes are shown in Section~\ref{sec:result} and we summarise our work and conclusions in Section~\ref{sec:summary}.

\section{Observations and sample}
\label{sec:obs}
\subsection{Observation, reduction and analysis}
\label{sec:ora}
Mapping Nearby Galaxies at Apache Point Observatory~\citep[MaNGA;][]{2015ApJ...798....7B} is an integral-field spectroscopic survey, part of the fourth-generation Sloan Digital Sky Survey \citep[SDSS-IV;][]{2017AJ....154...28B}. 
MaNGA employs $17$ Integral Field Units (IFUs), with each IFU consisting of $19$ to $127$ fibres arranged in a hexagonal bundle \citep{2015AJ....149...77D}. MaNGA makes use of the BOSS spectrographs~\citep{2013AJ....146...32S} on the 2.5m Sloan Telescope \citep{2006AJ....131.2332G}, covering a wavelength range of $3600 - 10000$ \AA\ with a spectral resolution of $60$ km/s (instrumental dispersion). The MaNGA galaxy sample is selected such that the galaxies have a spatial coverage of 1.5 (primary sample) or 2.5 (secondary) effective radii ($R_\mathrm{e}$), while maintaining a flat sample distribution in $i$-band absolute magnitude~\citep{2017AJ....154...86W}.
The MaNGA observing strategy is described in~\citet{2015AJ....150...19L}, while the spectrophotometric calibration strategy can be found in \citet{2016AJ....151....8Y}. \citet{2016AJ....152...83L} describes the MaNGA data reduction pipeline or DRP.

The MaNGA data analysis pipeline (DAP) processes and analyses the data cubes generated by the DRP, and delivers maps of stellar and gas properties for the MaNGA galaxies~\citep{2019AJ....158..231W}. The DAP obtains the stellar kinematics that we will use in this paper by fitting the observed spectra with the penalized pixel-fitting (pPXF) method~\citep{2004PASP..116..138C}, using the MILES stellar template library~\citep{2006MNRAS.371..703S,2011A&A...532A..95F} as templates. The gaseous kinematics are fitted simultaneously with pPXF using constructed emission line templates. 

The MaNGA data and maps used in this work were released in data release 15 \citep[DR15;][]{2019ApJS..240...23A} and we work with the Voronoi binned~\citep{2003MNRAS.342..345C} stellar and gas kinematics. The Voronoi binned spectra have a minimum signal-to-noise ratio $S/N = 10$, to ensure that the kinematic properties are accurately measured. In this work, we mainly use the line-of-sight mean velocity and line-of-sight velocity dispersion maps for the stellar kinematics, and the $\rm H\alpha$ emission line velocities for the gas kinematics. 

We note that the line-spread function (LSF) in MaNGA DR15 (and in previous releases) is underestimated, as determined in the recent internal MaNGA data release MPL-10~\citep[see][]{2021AJ....161...52L}, and will be corrected in future data releases. This leads to the systematic overestimation of the velocity dispersion, and we discuss this LSF effect to the results in the corresponding sections of this paper. 

\subsection{Sample selection}
\label{sec:sample}
Our parent sample consists of galaxies from the MaNGA DR15 sample that also have a classification in Galaxy Zoo 2 \citep{2013MNRAS.435.2835W,2016MNRAS.461.3663H}. The overlap between DR15 and Galaxy Zoo 2 is $3841$ galaxies. 

\begin{figure}
	\includegraphics[width=\columnwidth]{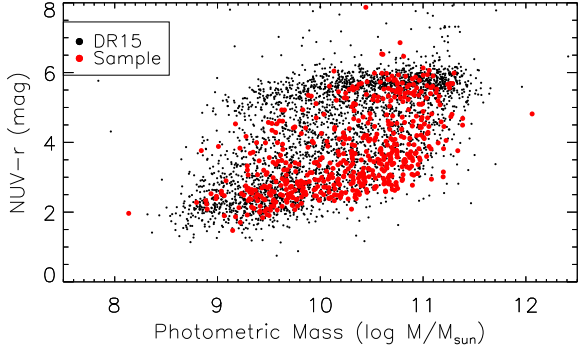}
    \caption{The colour-mass relation of the initial sample (red dots, $558$ galaxies) compared to the DR15 parent sample with Galaxy Zoo 2 classification (black dots, $3841$ galaxies). The NUV-$r$ colour and the photometric mass are obtained from the NSA catalogue~\citep{2011AJ....142...31B}.}
    \label{fig:colormass}
\end{figure}

In order to apply the analytical model, the galaxies in this sample are required to have the following features: i) regular morphologies; ii) fairly face-on (to allow for modelling the velocity and velocity dispersion fields together); iii) regular kinematics, including aligned stellar and gaseous velocity field. For our first criterion, we select galaxies without odd features in their optical images according to Galaxy Zoo 2 (as captured by the requirement t06\_odd\_a15\_no\_flag equals 1.0). We then pick reasonably face-on galaxies by requiring S\'ersic profile axis ratios between $0.5$ and $0.996$ as reported by the NASA Sloan Atlas (NSA) catalogue~\citep{2011AJ....142...31B}, within where we find that the fitting of the kinematic fields measured by reduced-$\chi^2$ (see Section~\ref{sec:da} for definition) has no correlation with the axis ratio. Finally, we visually inspect the kinematic maps of the galaxies, and remove those with misaligned stellar and $\rm H\alpha$ velocity fields or other irregular kinematic features. 
The colour-mass map of the initial sample at this stage ($558$ galaxies) is shown in Figure~\ref{fig:colormass}, compared to the DR15 parent sample. The sample contains both red and blue galaxies, as well as a number of green valley galaxies, representative of the colour and mass distributions of the DR15 sample. 

Accurate position angles (PA) and inclinations ($i$) are important for our analytical model. We use the package \texttt{KINEMETRY} \citep{2006MNRAS.366..787K} to measure these quantities from the stellar velocity maps. We use the velocity maps instead of imaging, as in this way we directly trace the stellar disc and minimise the influence of e.g. the thickness of the stellar disc or other galaxy components. \texttt{KINEMETRY} generates best-fitting ellipses on the velocity maps by analysing the high-order moments of harmonic expansions along every ellipse. The kinematic PA and axis ratio Q at each radius are recorded, and the inclination of the stellar disc can be obtained using $\mathrm{Q} = \cos{i}$.

\begin{figure}
	\includegraphics[width=\columnwidth]{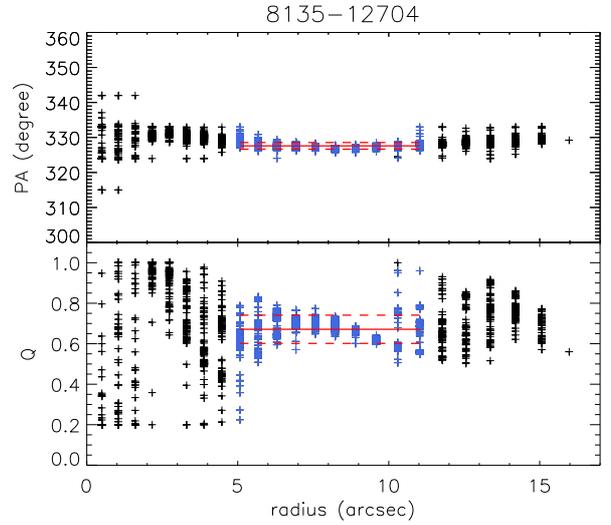}
    \caption{The distribution of PAs and Qs on each ellipse of an example galaxy (8135-12704). Blue plus signs stand for the values on the `good' ellipses which are included in the average, and the black plus signs show the values on the discarded ellipses. The red solid and dashed lines show the PA and Q of the galaxy and their 1-$\sigma$ uncertainties. We exclude the radii at which $\Delta \mathrm{Q} > 0.1$ or $\mathrm{Q_{med}} = 0.2$ or $1.0$.}
    \label{fig:paq}
\end{figure}

We adopt the following method to measure the kinematic PA and Q, and their uncertainties, for each galaxy. We perturb the stellar velocity maps 100 times with random Gaussian noise and fit the ellipses at the same radial positions for all perturbed maps with \texttt{KINEMETRY}.
A typical distribution of PAs and Qs of a sample galaxy is shown in Figure~\ref{fig:paq}.

\begin{figure}
	\includegraphics[width=\columnwidth]{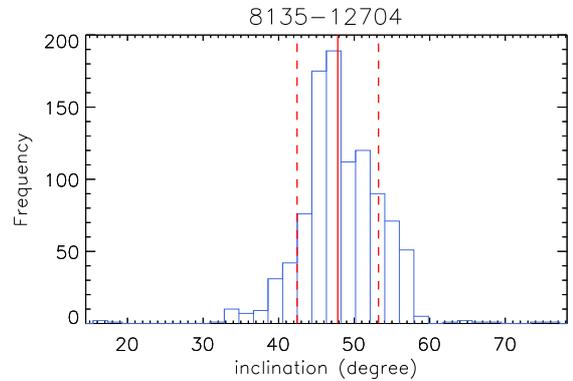}
    \caption{The distribution of the `effective' inclinations of example galaxy 8135-12704 (defined as the inclinations on the `good' ellipses without hitting the fitting boundaries). The red solid and dashed lines shows the inclination and 1-$\sigma$ uncertainties. }
    \label{fig:restincl}
\end{figure}

\begin{figure}
	\includegraphics[width=\columnwidth]{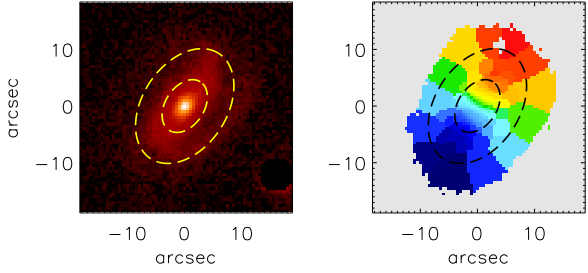}
    \caption{The `good' range of example galaxy 8135-12704 overlaid with its SDSS $r$-band image (left panel) and velocity field (right panel). The region between the dashed lines in each panel marks the radial range where we measure PA and Q.}
    \label{fig:range}
\end{figure}

Qs are often difficult to constrain or hit the fitting boundaries both in the very centre and in the outskirts of galaxies. In the centre of the galaxy their measurement can be affected by bulges or bars, or be hindered because of the limited amount of data points on the smaller ellipses. In the outskirts, the Voronoi bins are usually large due to the individual spectra having lower $S/N$. Therefore, we adopt the following method to obtain a robust inclination for each galaxy. We first empirically exclude the ellipses with a standard deviation of Qs ($\Delta \mathrm{Q}$) higher than $0.1$ or the ellipses at which the median value of Qs ($\mathrm{Q_{med}}$) hits the allowed limits ($[0.2, 1.0]$). The remaining ellipses are regarded as `good' ellipses. We then exclude the outliers and the measurements that are hitting the fitting boundaries on these `good' ellipses, leaving us with a smaller number of reliable `effective' data points. We then take the average of these `effective' data points on the `good' ellipses to obtain the PA and the inclination of the galaxy, as marked in blue in Figure~\ref{fig:paq}. As an example, the inclination distribution of galaxy 8135-12704 after converting the obtained Qs is shown in Figure~\ref{fig:restincl}. 
We mark the radial `good' range in its SDSS $r$-band image and velocity field for galaxy 8135-12704, as shown in Figure~\ref{fig:range}. The central region affected by a bulge or bar and the outskirts with a large bin size are thus excluded in the measurements of PA and Q.

Some galaxies display a uniform inclination distribution, and therefore we cannot decide a best-fitting inclination: these galaxies are excluded from the sample. We finally are left with a reliable best-fitting PA and inclination for $465$ galaxies.

\section{Methods}
\label{sec:methods}
In this section, we first introduce our analytical model, including the descriptions of the stellar velocity and dispersion profiles, the asymmetric drift correction of the stellar velocity based on a thin-disc assumption and the dynamical mass densities. We then introduce the data analysis procedures and our classification scheme based on our model.

\subsection{Models}

\subsubsection{Stellar velocity and dispersion}
We assume that the stellar azimuthal velocity $v_\phi$ follows the Evans model~\citep{1993MNRAS.260..191E,1994MNRAS.271..202E}, which considers the odd parts of the distribution function to describe the distribution of velocities in a flattened spherical system: 
\begin{equation}
\label{equ:vphi}
v_\phi = \frac{A R}{(R_\mathrm{c}^2+R^2+z^2/q_\phi^2)^{1/2+\delta/4}},
\end{equation} 
where $A$ is a constant, $R_\mathrm{c}$ is the core radius, $\delta$ stands for the logarithm slope of the rotation curve, and $q_\phi$ reflects the flattening of the equipotentials. $R$, $\phi$ and $z$ are used in their usual capacity as cylindrical coordinates.

Because the field of view for MaNGA is limited to 1.5 or 2.5 $R_\mathrm{e}$ at most, we cannot determine the logarithmic slope of the rotation curve $\delta$ from these data. We therefore simplify the model by assuming a flat rotation curve with $\delta = 0$ and under this assumption $A$ equals the maximal rotation velocity $v_\mathrm{max}$. In the disc plane ($z = 0$), the stellar velocity reduces to
\begin{equation}
\label{equ:vphi2}
v_\phi = \frac{v_\mathrm{max} \cdot R}{\sqrt{R^2 + R_\mathrm{c}^2}}.
\end{equation}
The line-of-sight velocity field is given by
\begin{equation}
\label{equ:vlos}
v_\mathrm{los}(R,\phi)= v_\phi(R)\cos\phi\sin i.
\end{equation}

We build the velocity dispersion model with an anisotropic component $\sigma_\mathrm{d}$ and an isotropic component $\sigma_\mathrm{iso}$.
We assume the anisotropic component follows an exponential profile of $\sigma_{R,\mathrm{d}} = \sigma_\mathrm{0,d}\exp(-R/h_{\sigma,\mathrm{d}})$~\citep{1970ApJ...160..811F}, describing the rotation-dominated component under a thin-disc assumption. 
The line-of-sight velocity dispersion will have the following form under the Evans model~\citep{2008MNRAS.383.1343W}:
\begin{equation}
    \label{equ:sigmaR}
    \sigma_\mathrm{los,d}^2 = \sigma_{R,\mathrm{d}}^2\left[ 1- \frac{R^2 \cos^2i}{\kappa q_\phi^2 (R^2 + 2R_\mathrm{c}^2)+R^2} -\frac{R^2 \sin^2i\cos^2\phi}{2(R^2 + R_\mathrm{c}^2)}\right],
\end{equation}
where for the thin-disc assumption with $z << R$, $\kappa$ is defined as:
\begin{equation}
\frac{\partial \overline{v_Rv_z}}{\partial z} = \kappa \frac{(\sigma_{R,\mathrm{d}}^2-\sigma_{z,\mathrm{d}}^2)}{R}.
\end{equation}
$\kappa$ is varied between 0 and 1, from the alignment of the velocity ellipsoid with the cylinder coordinate system to the spherical one. The gravitational potential $\Phi$ of the galaxy is not directly known, but if we assume the circular velocity $V_c \simeq v_\phi$, then the gravitational potential from the Evans model becomes:
\begin{equation}
    \Phi \simeq \frac{v_\mathrm{max}^2}{2} \ln{(R_\mathrm{c}^2+R^2+z^2/q_\phi^2)},
\end{equation}
According to~\citet{1991ApJ...368...79A}, 
\begin{equation}
    \kappa = \frac{R^2\Phi_{,Rzz}}{3\Phi_{,R}+R\Phi_{,RR}-4R\Phi_{,zz}} \Bigg|_{z=0},
    \label{morequ:kappa}
\end{equation}
where $\Phi_{,R} = \partial \Phi/ \partial R $, $\Phi_{,RR} = \partial^2 \Phi/ \partial R^2 $, etc.
We then obtain that $\kappa$ is a function of $q_\phi^2$ by
\begin{equation}
    \kappa \simeq \frac{R^2}{(2-q_\phi^2)R^2+2R_\mathrm{c}^2(1-q_\phi^2)}.
    \label{equ:kappa}
\end{equation}
$\kappa$ and $q_\phi$ are both related to the flattening of galaxies. For a spherical case, we have $\kappa = 1$ and $q_\phi = 1$, while for a typical disc $q_\phi \simeq 0.7$ and $\kappa \simeq 0.6$~\citep{1991ApJ...368...79A}. Since $\kappa$ and $q_\phi$ always appear in the form of $\kappa q_\phi^2$ in our model, we parameterise $\kappa$ and $q_\phi$ together empirically by
\begin{equation}
    \label{equ:t}
    \kappa q_\phi^2 = (1-t)\exp(-R/h_t) + t,
\end{equation}
where $ 0.1 \leqslant t \leqslant 1 $, which allows us to fit velocity profiles with both flattening cores and steep cusps in the centre of a galaxy.
The vertical velocity dispersion of the rotation-dominated component $\sigma_{z,\mathrm{d}}$ is then given in our model by
\begin{equation}
\label{equ:sigma_z}
\sigma_{z,\mathrm{d}}^2/\sigma_{R,\mathrm{d}}^2 = \left[1 - \frac{R^2}{\kappa q_\phi^2 (R^2 + 2 R_\mathrm{c}^2) + R^2} \right].
\end{equation}

We introduce the isotropic component to describe the velocity dispersion field which is not included in the thin disc component. We assume it follows an exponential profile $\sigma_\mathrm{los,iso} = \sigma_\mathrm{0,iso}\exp(-R/h_{\sigma,\mathrm{iso}})$. The line-of-sight velocity dispersion then becomes:
\begin{equation}
    \label{equ:sigmaLOS}
    \sigma_\mathrm{los}^2 = \sigma_\mathrm{los,d}^2 + \sigma_\mathrm{los,iso}^2
\end{equation}

Here we emphasise that Equation~\ref{equ:sigmaLOS} is not a component decomposition method which can obtain velocity distribution profiles for different components. This isotropic component is merely introduced to guarantee the fitting of the disc component, and we do not link this isotropic component to any one or multiple physical components (e.g. bulge, bar, nuclear disc/ring and AGN effect, etc).

\subsubsection{Asymmetric drift correction}
The asymmetric drift is the difference between the stellar velocity $v_\phi$ and the circular velocity $V_c$, which is approximated by the velocity of $\rm H\alpha$ in our model. If we correct the asymmetric drift for $v_\phi$ under a thin-disc assumption, we will obtain $V_c$ correctly only for the galaxies whose rotational velocity has a contribution from a thin disc. This then provides a kinematic way to identify thin discs in galaxies.

The asymmetric drift correction has the following form under a thin-disc assumption in the Evans model~\citep{2008gady.book.....B,2008MNRAS.383.1343W},
\begin{equation}
\begin{split}
       v_\mathrm{corr}^2 & = v_\phi ^2 - \sigma_{R,\mathrm{d}}^2 \left[ \frac{\partial \ln \mu_\mathrm{d}}{\partial \ln R} + \frac{\partial \ln \sigma_{R,\mathrm{d}}^2}{\partial \ln R} + \frac{R^2}{2(R^2 + R_\mathrm{c}^2)} \right. \\
       & + \left. \frac{\kappa R^2}{\kappa q_\phi ^2(R^2 + 2R_\mathrm{c}^2)+ R^2}\right].
\end{split}
\end{equation}
Here we only adopt the velocity dispersion of the anisotropic component $\sigma_\mathrm{d}^2$ for asymmetric drift correction, while the isotropic component $\sigma_\mathrm{iso}^2$, which is introduced only for the fitting of velocity dispersion and in our model is taken to represent the non-rotation component with low tangential velocities, is not included in the thin-disc model and therefore ignored in the correction. We also assume the flux density of the disc $\mu_\mathrm{d}$ follows an exponential profile with $\mu_\mathrm{d} \propto \sigma_{R,\mathrm{d}}^2$~\citep{2008gady.book.....B} and substitute $\kappa$ (see Equation~\ref{equ:kappa}), resulting in:
\begin{equation}
v_\mathrm{corr}^2 \simeq v_\phi ^2 - \sigma_{R,\mathrm{d}}^2 \left( \frac{2\partial \ln \sigma_{R,\mathrm{d}}^2}{\partial \ln R} + \frac{R^2}{R^2 + R_\mathrm{c}^2} \right).
\end{equation}

\subsubsection{Dynamical mass density}
The dynamical mass density of an isothermal thin disc is shown by~\citet{2008gady.book.....B} to be:
\begin{equation}
\Sigma_\mathrm{dyn,d} = \frac{\sigma_z^2}{k\pi G h_z},
\end{equation}
which is a function of vertical velocity dispersion $\sigma_z$ and disc scale height $h_z$. $k$ is a parameter that indicates different vertical mass distributions, and we adopt $k = 1.5$ as approximate for an exponential one~\citep{2010ApJ...716..234B}. $\sigma_z$ is derived in our velocity dispersion model following equation~\ref{equ:sigma_z}, and we only need to obtain $h_z$.

Due to the low inclinations of the sample galaxies, we need to estimate the disc scale heights from scaling relations. We adopt two independent scaling relations and take a weighted average of them. The first relation is the scale height versus the rotation velocity~\citep{2011ARA&A..49..301V}, 
\begin{equation}
h_{z,1} = (0.45\pm 0.05) (V_\mathrm{rot}/100 \mathrm{km/s}) - (0.14 \pm 0.07)\mathrm{kpc}.
\end{equation}
We take the maximal velocity $v_\mathrm{max}$ of the Evans model as the rotation velocity.
The second relation is the scale height versus the scale length~\citep{2010ApJ...716..234B},
\begin{equation}
\label{equation:SH}
\log(h_\mu/h_{z,2}) = 0.367 \log(h_\mu/\mathrm{kpc}) + 0.708 \pm 0.095.
\end{equation}
We adopt $h_\mu = 0.5 h_{\sigma,\mathrm{d}}$~\citep{2013A&A...557A.131M}, with the latter obtained from the fitting of the velocity dispersion profile.
We then take the weighted average of $h_{z,1}$ and $h_{z,2}$ as the scale height $h_z$. We also use the intrinsic scatter of these two scaling relations to estimate the uncertainties of $h_{z,1}$ and $h_{z,2}$, and then to obtain the uncertainty of $h_z$.

\subsection{Data analysis}
\label{sec:da}
There are 8 free parameters directly fitted from the data in our thin-disc model, which are summarised as follows. The maximum stellar velocity $v_\mathrm{max}$ and core radius $R_\mathrm{c}$ fitted to the stellar velocity field according to Equation~\ref{equ:vphi2} \&~\ref{equ:vlos}; The parameters regarding to the stellar velocity dispersion ($\sigma_\mathrm{0,iso}$ and $h_{\sigma,\mathrm{iso}}$ for the exponential isotropic component, $\sigma_\mathrm{0,d}$ and $h_{\sigma,\mathrm{d}}$ for the exponential radial profile of the disc component) and the flattening of the velocity ellipsoid (parameterised as $t$ and $h_t$ using an exponential profile) are fitted to the stellar velocity dispersion field according to Equation~\ref{equ:sigmaR},~\ref{equ:t} \&~\ref{equ:sigmaLOS}. These parameters and their corresponding statistical uncertainties are obtained with the non-linear least-squares fitting program \texttt{MPFIT}~\citep{2009ASPC..411..251M}.

We convolve the modelled kinematic maps with the MaNGA PSF before comparing with observations. Following the procedure as outlined in~\citet{1989A&A...223...47B}, we construct the convolved models by convolving a kernel which takes the seeing of each galaxy and the spatial sampling of the kinematic maps (0.5 arcsec) into consideration. This method is equivalent to convolving a Gaussian kernel which takes the size of the reconstructed MaNGA PSF~\citep[FWHM = 2.5 arcsec,][]{2016AJ....152...83L}.

The goodness of the fitting of the kinematic map (velocity or velocity dispersion) is measured by its reduced-$\chi^2$, defined as
\begin{equation}
    \chi^2/\mathrm{dof} = \left[\sum \left(\frac{u_\mathrm{obs}-u_\mathrm{mod}}{e_\mathrm{obs}}\right)^2\right]/\mathrm{dof},
\end{equation}
where $u_\mathrm{obs}$ stands for the observed data and $e_\mathrm{obs}$ stands for the uncertainties of the observed data, $u_\mathrm{mod}$ stands for the modelled data, and $\mathrm{dof}$ is the degree of freedom for which we adopt the number of data points as an approximation. We obtain the reduced-$\chi^2$ for both velocity and velocity dispersion to measure whether the kinematic map is well fitted.

We have built our thin-disc model to describe galaxy kinematic properties with two major assumptions: 
\begin{enumerate}
    \item The stellar velocity corrected for asymmetric drift matches the $\rm H\alpha$ velocity for galaxies with a thin disc.
    \item The velocity dispersion is composed of a disc component and an isotropic component, the former taking a dominant role in the galaxy and the influence of the latter being negligible for a disc-dominated galaxy.
\end{enumerate}
We examine whether these assumptions are fulfilled by building corresponding criteria.

The case that assumption (i) is not fulfilled indicates our model detects no thin disc features in the galaxy. Considering the strong non-circular motion and low S/N data in the gaseous velocity field, it is difficult to measure the similarity of the corrected stellar velocity field and the gaseous velocity field as a reduced-$\chi^2$. We therefore quantify the difference between these two velocity fields by introducing a parameter $\zeta$, defined as the median of the relative residuals of the map:
\begin{equation}
\label{equ:zeta}
    \zeta = \mathrm{median}(|(v_\mathrm{gas}-v_\mathrm{corr})/v_\mathrm{gas}|).
\end{equation}

For the rest of the galaxies with a thin disc, we need to check if assumption (ii) is fulfilled, which demands the thin disc taking a dominant role in the galaxy by guaranteeing the effect of the isotropic component of the dispersion is insignificant. We introduce a parameter $\eta$, which accounts for both the magnitude and influence area of the dispersion components:
\begin{equation}
\label{equ:eta}
    \eta = {\sigma_\mathrm{0,iso} h_{\sigma,\mathrm{iso}}^2 }/ {\sigma_\mathrm{0,d} h_{\sigma,\mathrm{d}}^2}.
\end{equation}
A high $\eta$ value indicates the isotropic component in the galaxy takes a dominant role.

\section{Results}
\label{sec:modelfitness}

In this section, we introduce the results of applying our model to the sample following the methods described in Section~\ref{sec:methods}.

\begin{figure}
	\includegraphics[width=0.9\columnwidth]{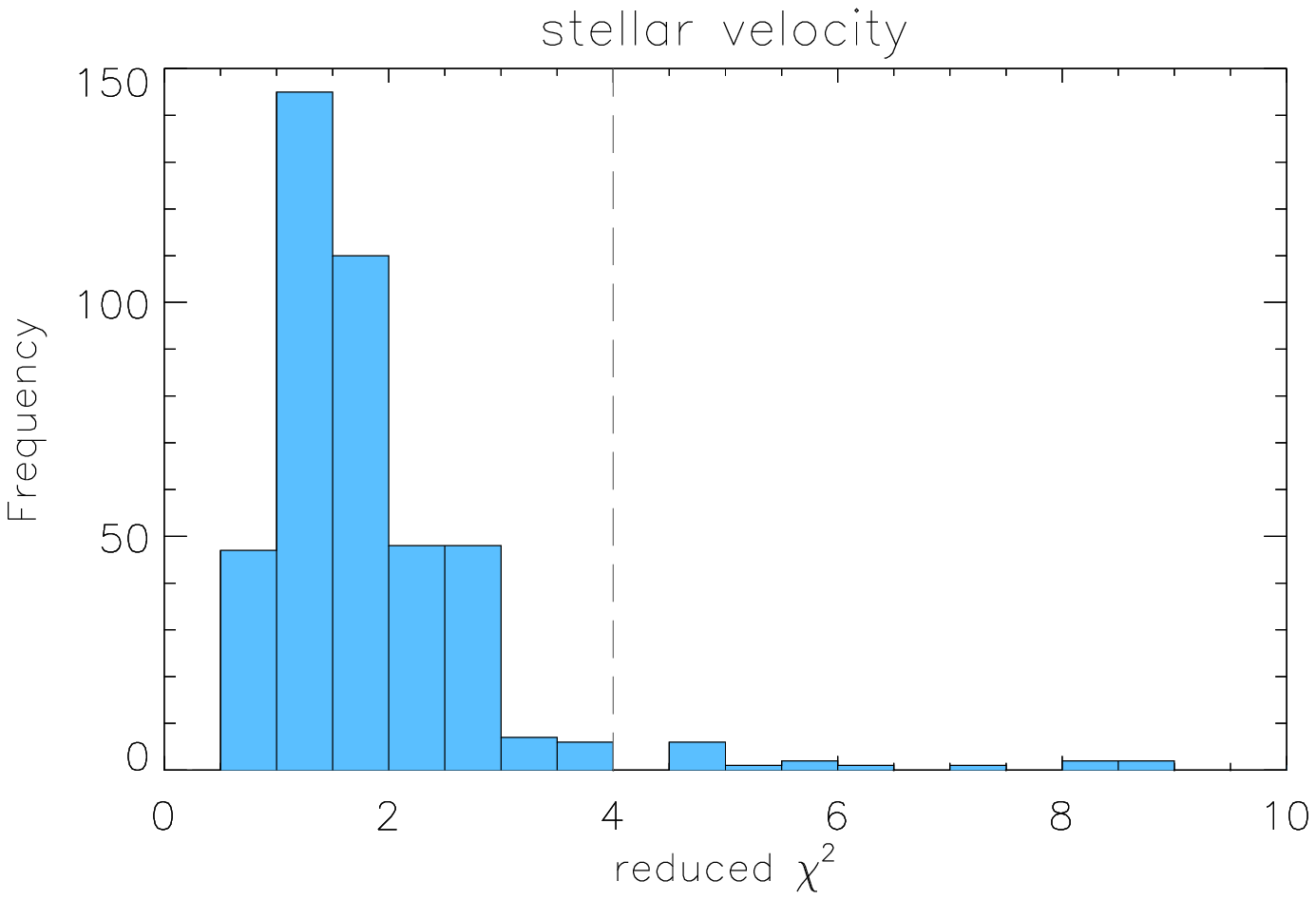}
	\includegraphics[width=0.9\columnwidth]{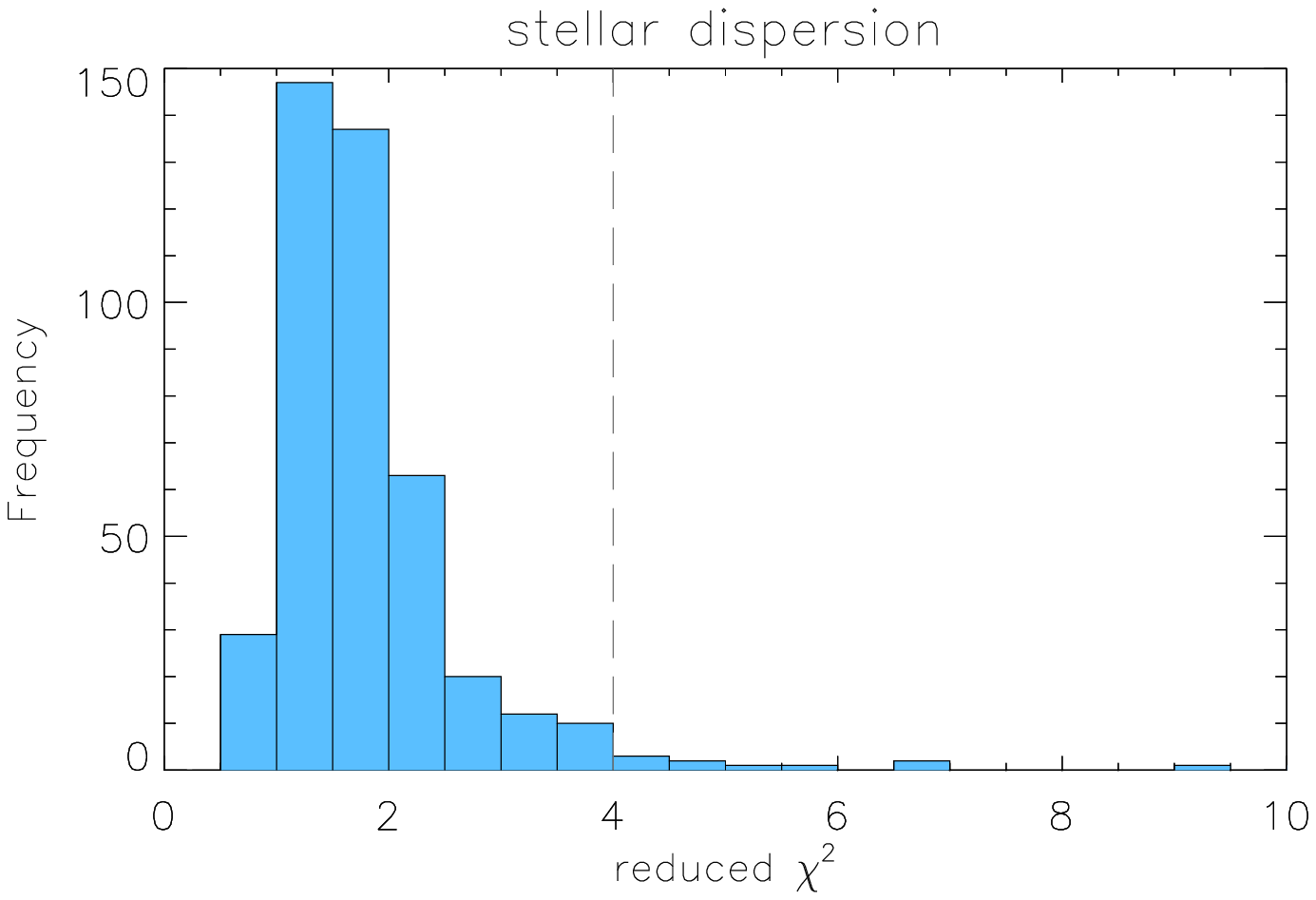}
    \caption{Reduced-$\chi^2$ distribution of the stellar velocity (upper panel) and the stellar velocity dispersion (lower panel) for the full sample of 430 galaxies. The galaxies right of the dashed lines are regarded as poorly modelled and are excluded from further analysis. }  
    \label{fig:distribute}
\end{figure}

\begin{figure}
	\includegraphics[width=\columnwidth]{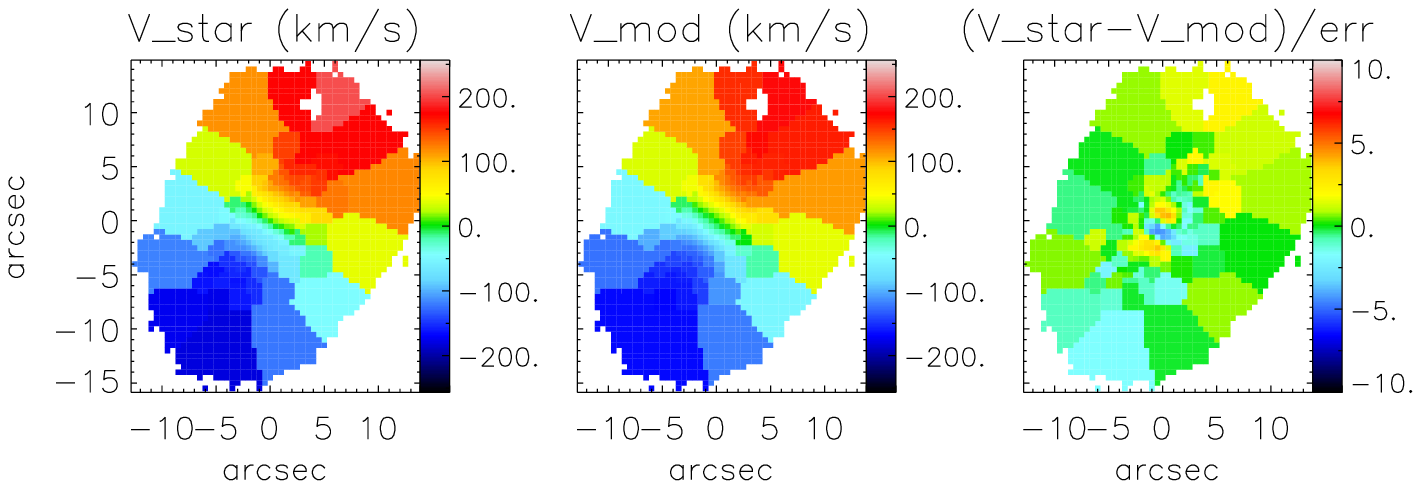}
	\includegraphics[width=\columnwidth]{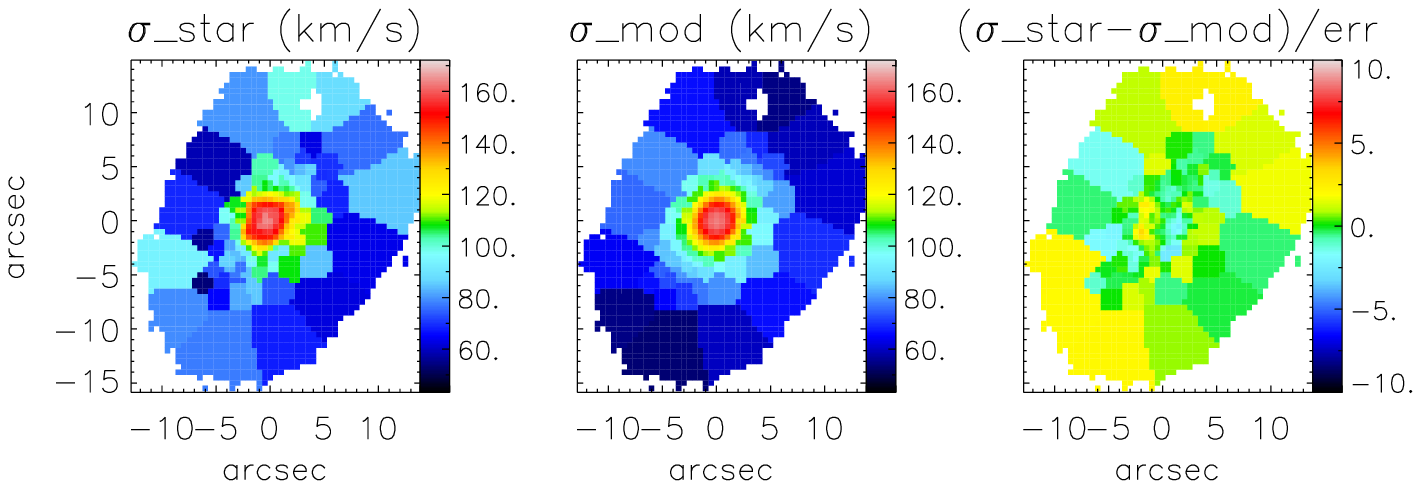}
    \caption{Stellar velocity (top) and velocity dispersion (bottom) fitting of galaxy 8135-12704. From left to right: data, model and residuals. This galaxy is well represented by our analytical thin disc model. }
    \label{fig:starmodel}
\end{figure}

The kinematic fitting of the observed kinematic maps with this model converged for 430 galaxies, and we show the histograms of reduced-$\chi^2$ of the stellar velocity and the stellar dispersion in Figure~\ref{fig:distribute}.
We subsequently make a cutoff at reduced-$\chi^2 = 4.0$ for both velocity and velocity dispersion empirically to exclude the galaxies which are poorly modelled. We also exclude galaxies with thin-disc parameters poorly fitted i.e. $\sigma_\mathrm{0,d}$ has an uncertainty higher than 0.5 dex or $h_{\sigma,\mathrm{d}}$ has an an uncertainty higher than 1 dex in logarithmic scale, which will result in the dynamical mass density in logarithmic scale have an uncertainty higher than 1 dex. The parameters of the isotropic component $\sigma_\mathrm{0,iso}$ and $h_{\sigma,\mathrm{iso}}$ are not well constrained for some galaxies, but we do not exclude these galaxies because it has little impact on the results. After this, we have $389$ galaxies left in the sample. We show an example of the fitted models in Figure~\ref{fig:starmodel}.

\begin{figure}  
	\includegraphics[width=\columnwidth]{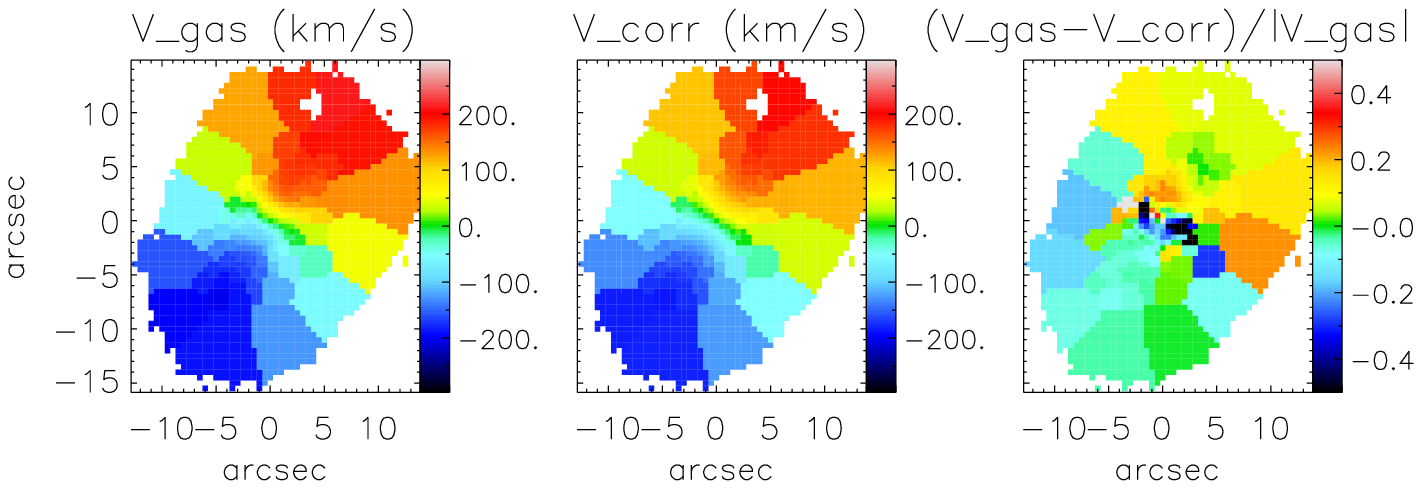}
	\includegraphics[width=\columnwidth]{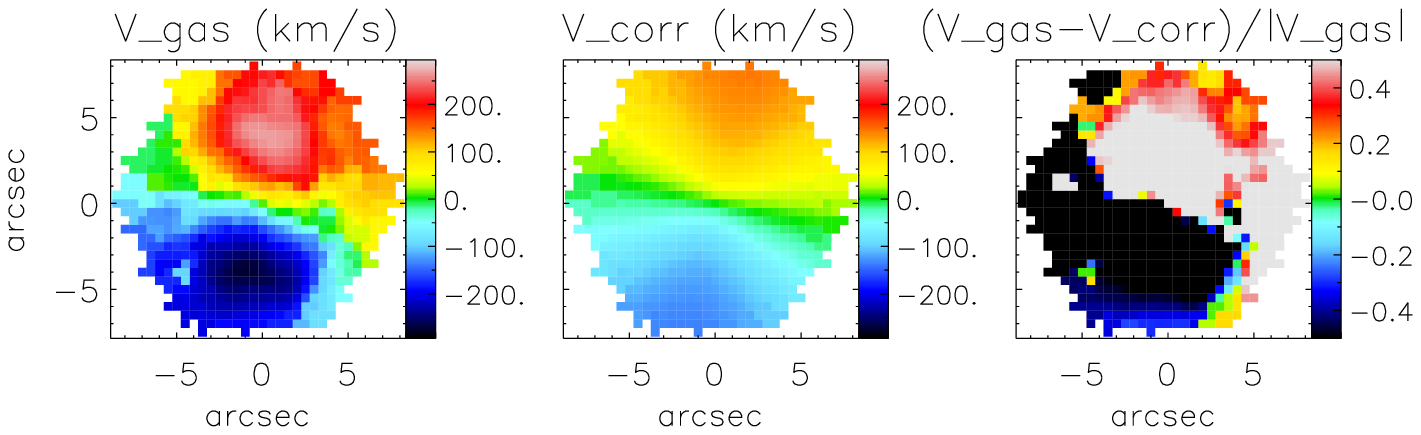}
    \caption{Comparison between $\rm H\alpha$ velocity and asymmetric drift corrected stellar velocity of galaxies 8135-12704 (top) and 7957-3701 (bottom). From left to right: $\rm H\alpha$ velocity, corrected stellar velocity and the relative difference. 8135-12704 is an example of a galaxy with matched corrected stellar velocity and $\rm H\alpha$ velocity, while 7495-3701 represents a failed case.}
\label{fig:starcorr}
\end{figure}

We compare the corrected stellar velocity for asymmetric drift with the $\rm H\alpha$ velocity representing the circular velocity for 378 out of our 389 galaxies which have regular $\rm H\alpha$ velocity fields in the outskirts as our final sample. For most galaxies, the corrected stellar velocity derived from our model is consistent with $\rm H\alpha$ velocity, while for some of galaxies, the corrected stellar velocity strongly deviates from the $\rm H\alpha$ velocity, indicating the failure of our thin-disc model. We show an example for each case in Figure~\ref{fig:starcorr}. 

\begin{figure}
	\includegraphics[width=\columnwidth]{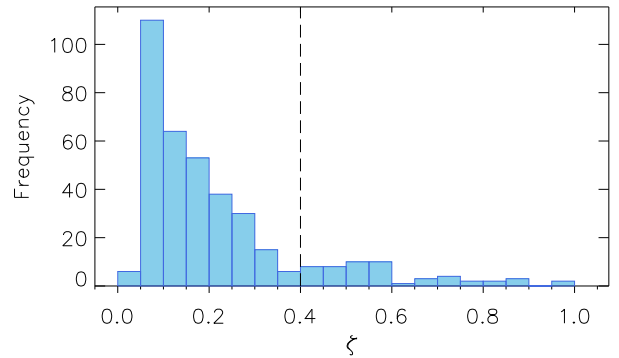}
    \caption{$\zeta$-distribution of the final sample (378 galaxies). The galaxies right to the dashed line are regarded as disc-free galaxies.}
    \label{fig:class}
\end{figure}

According to the distributions of $\zeta$ (defined in Equation~\ref{equ:zeta}) shown in Figure~\ref{fig:class} and based on visual inspection, we make a cut-off at $\zeta = 0.4$. Galaxies with $\zeta > 0.4$ are regarded as no detections of thin discs and therefore classified as disc-free galaxies.
We make a cutoff at $\eta = 1.0$ (defined in Equation~\ref{equ:eta}), where the two components have a comparable influence, and further divide the galaxies with a thin disc detection into disc-dominated galaxies ($\eta \leqslant 1.0$) and non-disc-dominated galaxies ($\eta > 1.0$). 

\begin{figure}
    \centering
    \includegraphics[width=\columnwidth]{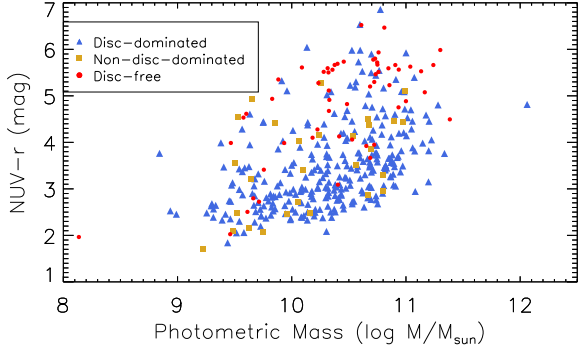}
    \caption{The colour-mass plane of the final sample (378 galaxies). Blue triangles stand for disc-dominated galaxies, yellow squares stand for non-disc-dominated galaxies and red circles stand for disc-free galaxies. We find that the disc-dominated galaxies can be found in both the blue cloud and red sequence, while the disc-free galaxies are mostly located in the latter.}
    \label{fig:sample-colour-mass}
\end{figure}

The colour-mass relation of the classifications is shown in Figure~\ref{fig:sample-colour-mass}, in which disc-free galaxies are predominantly located on the red sequence, while the disc-dominated galaxies show a larger variety in both colour and mass. We further explore the visual morphology of our kinematic classification in Section~\ref{subsec:morph}.

We examine the stability of the fitting of the velocity dispersion fields for a sub-sample of galaxies which are randomly selected from each kinematic type.
We perturb the velocity dispersion fields by adding Gaussian noise to the data, and the standard deviations of the Gaussian noise being the $1-\sigma$ uncertainties of observed velocity dispersion. For each galaxy in this sub-sample, we create 100 perturbed velocity dispersion fields and repeat the fitting.
The results show that our classification of galaxies is stable to the perturbation, therefore, the classification scheme based on the model is valid. 
In general, the model parameters related to the thin disc ($\sigma_\mathrm{0,d}$, $h_{\sigma,\mathrm{d}}$, $t$ and $h_t$) are stable to the perturbation for disc-dominated galaxies, with a typical uncertainty compatible to the statistical errors given by the fitting procedure (within 10\%). We also find a slight correlation between $\sigma_\mathrm{0,d}$ and $h_{\sigma,\mathrm{d}}$ around the best-fitting values such that  $h_{\sigma,\mathrm{d}}$ decreases as $\sigma_\mathrm{0,d}$ rises. The fitting of the isotropic component ($\sigma_\mathrm{0,iso}$ and $h_{\sigma,\mathrm{iso}}$) are less stable with larger uncertainties affects the values of $\eta$. Since the variations hardly cross the cut-off of $\eta = 1.0$, they have little effect on the classification. 

As we have mentioned in Section~\ref{sec:ora}, the LSF effect Section introduces a systematic overestimation of velocity dispersion in MaNGA DR 15, especially in the outskirts of galaxies~\citep[about 3\% at 80 km/s, about 10\% at 30-40 km/s, see ][]{2021AJ....161...52L}. This affects the fitting of velocity dispersion fields systematically with the disc parameters being overestimated (mainly $h_{\sigma,\mathrm{d}}$ being overestimated by no more than 10\%). This LSF effect has little impact on the classification.

We remind the readers that the velocity dispersion model is descriptive, therefore, we cannot link the isotropic component to any physical structure of the galaxy (e.g. bulge, bars and nuclear discs or rings, AGN effects, etc). As a consequence, we believe our thin-disc model is able to distinguish thin-disc features in galaxies in general and to study the overall trends of galaxies of different kinematic types, but is not completely reliable on the measurements of the model parameters in a specific individual galaxy.

\section{Discussions}
\label{sec:result}

\subsection{Stellar angular momentum}
\citet{2007MNRAS.379..401E} introduced $\lambda_R$ as a proxy of stellar angular momentum per unit mass, which correlates with the intrinsic morphology and dynamics of galaxies. According to their positions on the $\lambda_R-\epsilon$ diagram, early-type galaxies are classified as fast and slow rotators dominated by circular and random motions, respectively~\citep{2007MNRAS.379..401E,2007MNRAS.379..418C}, while late-type galaxies are expected to be dominated by circular motions~\citep{2019A&A...632A..59F}. 

\begin{figure}
    \includegraphics[width=\columnwidth]{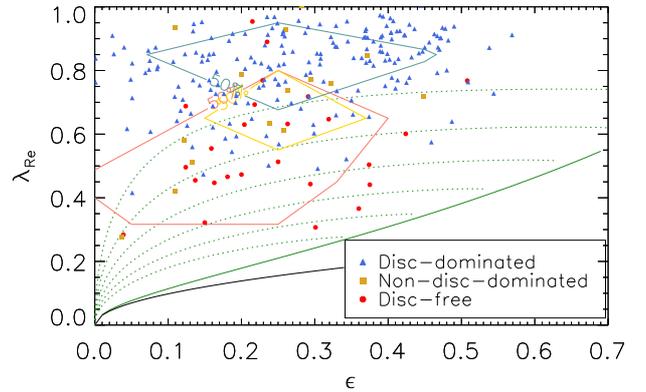}
    \caption{$\lambda_{R_\mathrm{e}}$ versus ellipticity $\epsilon$ for 230 galaxies in the sample. Blue triangles stand for disc-dominated galaxies, yellow squares stand for non-disc-dominated galaxies and red circles stand for disc-free galaxies. The contours plot half the maximum densities of each type. The black solid line $\lambda_{R_\mathrm{e}} = 0.31 \sqrt{\epsilon}$ distinguishes fast and slow rotators~\citep{2011MNRAS.414..888E}, and the green solid line corresponds with the theoretical curve of anisotropy $\delta = 0.7\epsilon_\mathrm{intr}$, where $\epsilon_\mathrm{intr}$ is the intrinsic ellipticity~\citep{2007MNRAS.379..418C}. The green dotted lines show the locations of galaxies with intrinsic ellipticity $\epsilon_\mathrm{intr}$ varied between $[0.35,0.95]$ in steps of $0.1$. The disc-dominated galaxies and disc-free galaxies are clearly distinct in the plot, and the non-disc-dominated galaxies lie in between these two types.}
    \label{fig:lambda-ell}
\end{figure}

The galaxy sample we worked with was selected with the requirement of regular rotation features, and hence consists of mostly late-type galaxies and the early-type fast rotators. To investigate whether our classification is able to distinguish intrinsic morphology, we adopt the spin parameter $\lambda_{R_\mathrm{e}}$ introduced as a proxy for the stellar angular momentum within the half-light ellipse, and ellipticity $\epsilon$ obtained from~\citet{2018MNRAS.477.4711G}. We plot the $\lambda_{R_\mathrm{e}} - \epsilon$ diagram showing our galaxies in common with their sample in Figure~\ref{fig:lambda-ell}.
The figure shows that disc-dominated galaxies and disc-free galaxies are clearly distinct in the $\lambda_{R_\mathrm{e}} - \epsilon$ diagram. The disc-dominated galaxies occupy the top region in the diagram ($\lambda_{R_\mathrm{e}} > 0.5$) while the bottom ($\lambda_{R_\mathrm{e}} < 0.5$) is occupied by the disc-free galaxies. The non-disc-dominated galaxies lie in between these regions. This distribution is consistent with the result obtained by~\citet{2020MNRAS.495.4638O}, which decomposes bulges and discs with stellar kinematics and find they occupy the bottom and top of the $\lambda_{R_\mathrm{e}} - \epsilon$ diagram, respectively. This result indicates that the thin-disc model is capable of discerning thin discs, which are correlated with higher $\lambda_{R_\mathrm{e}}$, and that our classification method reflects the different kinematic properties between the three different types.

\subsection{Mass property}
\label{subsec:dm}

Mass distribution is a fundamental property of galaxies and plays a crucial role in galaxy evolution. The dynamical mass density can provide constraints on the stellar mass density, thereby constraining stellar evolution models and the dark matter density. 

We obtain the disc dynamical mass density $\Sigma_\mathrm{dyn,d}$ with disc vertical velocity dispersion $\sigma_{z,\mathrm{d}}$ and scale height $h_z$, and obtain the corresponding uncertainty of $\Sigma_\mathrm{dyn,d}$ from the fitting error of $\sigma_{z,\mathrm{d}}$ and the uncertainty of $h_z$ using error propagation. We compare the disc dynamical mass density $\Sigma_\mathrm{dyn,d}$ with the stellar mass density in the DR15 MaNGA FIREFLY value added catalogue~\citep[][]{2017MNRAS.466.4731G}. The stellar mass density is obtained with the spectral fitting code \texttt{FIREFLY} using high spectral resolution stellar population models~\citep{2011MNRAS.418.2785M} and a Kroupa initial mass function ~\citep[IMF,][]{2001MNRAS.322..231K}. 

\begin{figure}
	\includegraphics[width=\columnwidth]{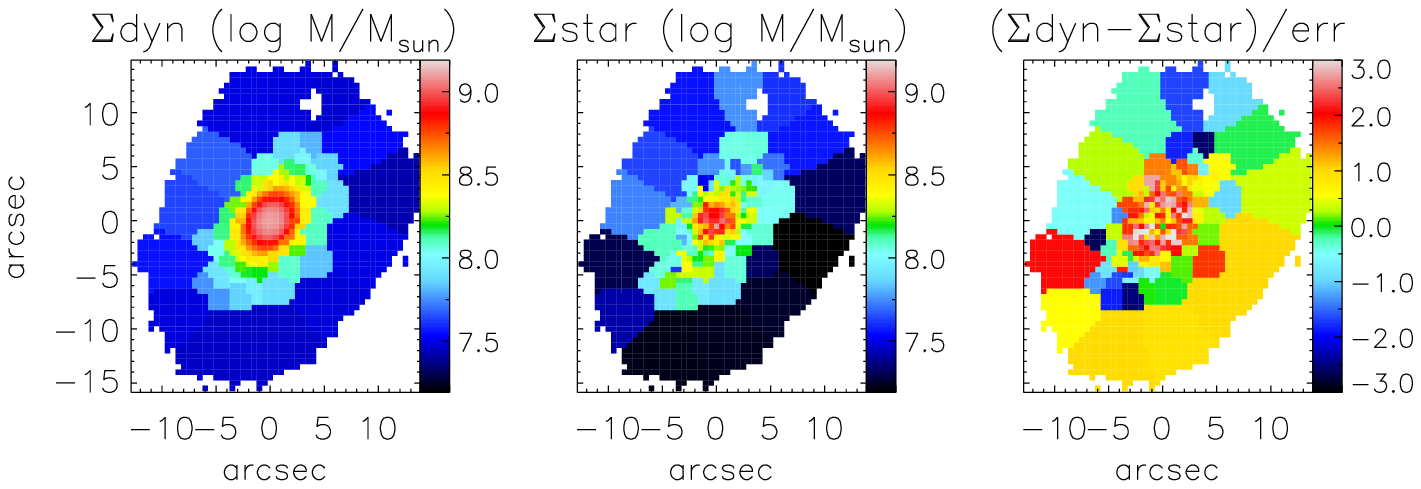}
	\includegraphics[width=\columnwidth]{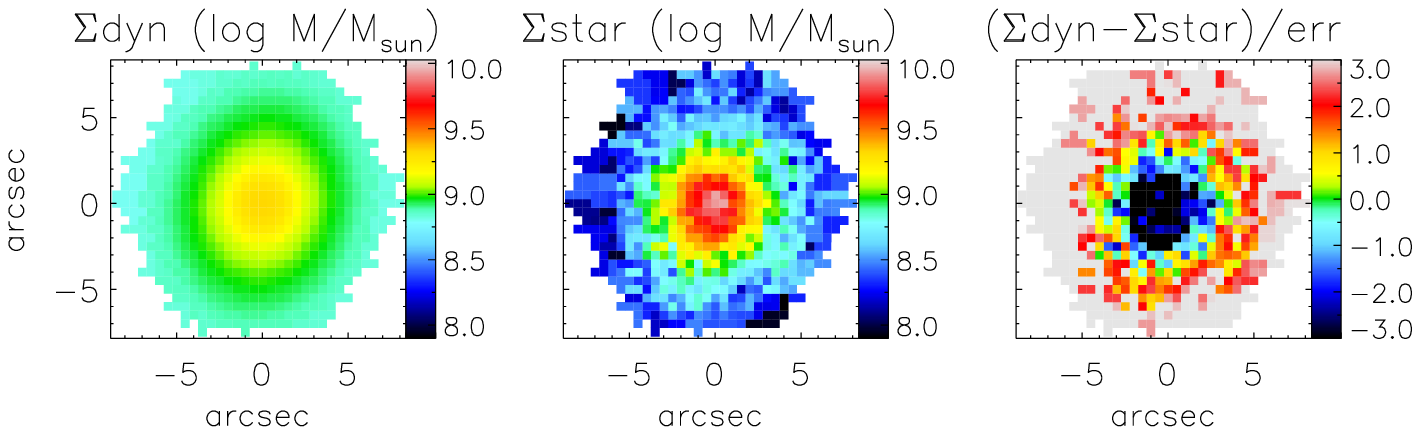}
    \caption{Comparison between dynamical and stellar mass densities ($\mathrm{M_\odot}^{-1}\mathrm{kpc}^{-2}$ in logarithmic scale) of a disc-dominated galaxy 8135-12704 (top) and a disc-free galaxy 7957-3701 (bottom). From left to right: dynamical mass density, stellar mass density and the relative difference.}
    \label{fig:dynmass}
\end{figure}

In Figure~\ref{fig:dynmass} we show a comparison between the disc dynamical and stellar mass densities, using the same two example galaxies shown in Figure~\ref{fig:starcorr}. 
For galaxy 8135-12704, the disc dynamical mass density is in agreement with the outskirts of the stellar mass density, indicating a small fraction of non-stellar matter. 
In the inner region, although the velocity dispersion field is fitted by parameterising $\kappa q_\phi^2$ and introducing an isotropic component, the thin-disc assumption in our model might be invalid because the possible existence of components other than the thin disc. Therefore, the disc dynamical mass density probably cannot trace the real mass distribution in the inner region. 
In contrast, the dynamical mass density of galaxy 7495-3701 differs from the stellar mass density throughout, mainly a result of the failure of the thin-disc model in Figure~\ref{fig:starcorr}. 

The comparison between disc dynamical mass density and stellar mass density for each galaxy is not always straightforward. To ease comparison, we calculate global parameters representing the disc dynamical mass and stellar mass within the MaNGA IFU, by integrating over the spaxels. We will use these parameters, $M_\mathrm{dyn}$ and $M_*$ in further discussions within this paper. The corresponding uncertainties of $M_\mathrm{dyn}$ and $M_*$ are calculated from the uncertainty of $\Sigma_\mathrm{dyn,d}$ and the uncertainty of stellar mass density in the DR15 MaNGA FIREFLY value added catalogue, respectively.

\begin{figure}
    \centering
    \includegraphics[width=\columnwidth]{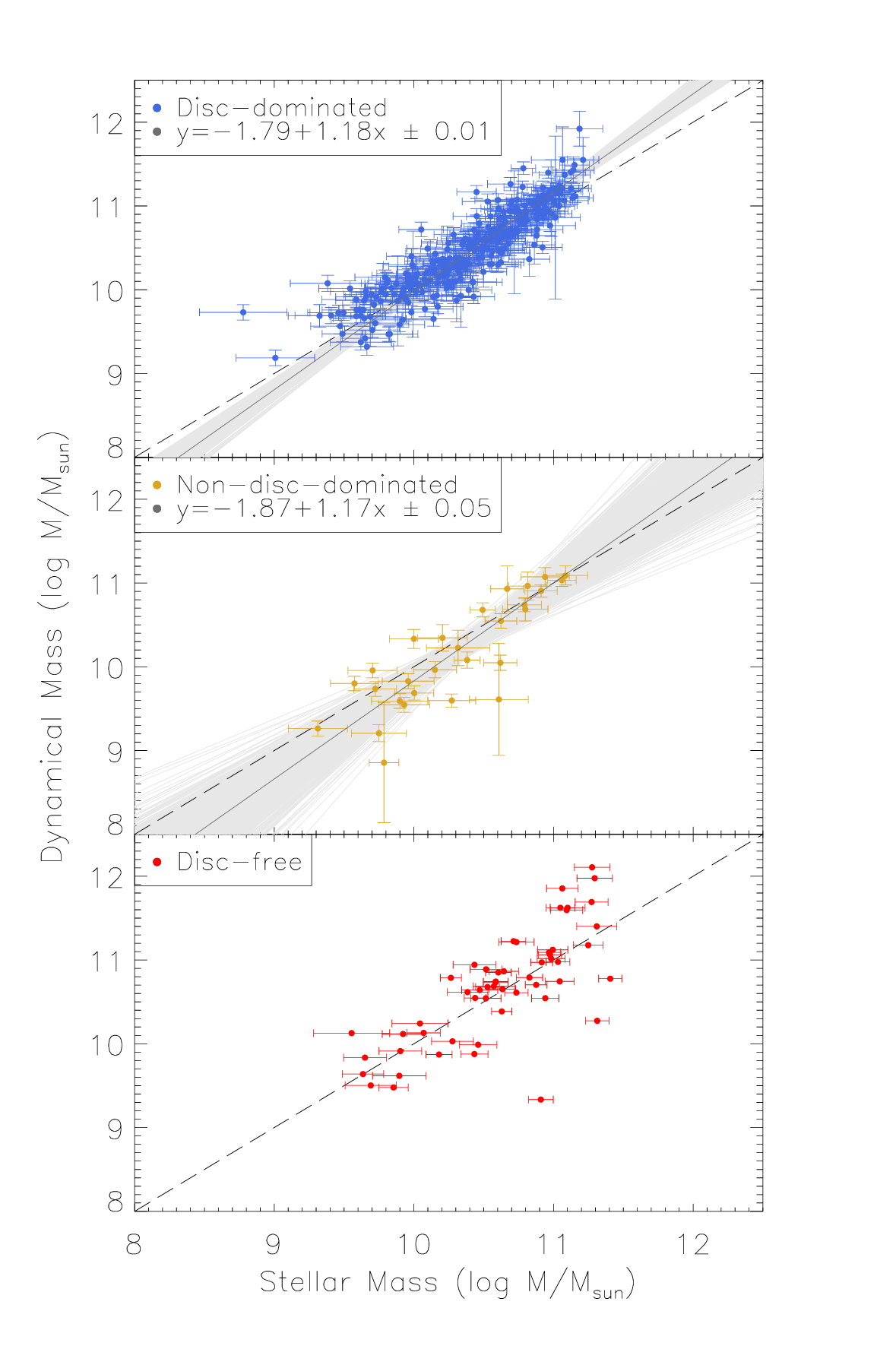}
    \caption{Dynamical mass versus stellar mass (in logarithmic scale) within MaNGA coverage. From top to bottom: disc-dominated galaxies, non-disc-dominated galaxies and disc-free galaxies. For disc-dominated and non-disc-dominated galaxies, we also fit a linear regression to the logarithmic stellar and disc dynamical mass as shown in the legends (the dark grey solid line), and light grey lines are the corresponding uncertainty range including the uncertainties of regression coefficients (slope and intercept). The black dashed lines in each panel are the 1-to-1 lines.}
    \label{fig:allmass}
\end{figure}

We show the $M_\mathrm{dyn}-M_*$ correlation of each galaxy type in Figure~\ref{fig:allmass}.
In general, the data points in each panel are distributed around the 1-to-1 line in the figure. The dynamical mass and stellar mass of disc-dominated galaxies are well correlated, while the correlation for non-disc-dominated galaxies has a larger scatter and several outliers. There is no clear correlation for disc-free galaxies as the intrinsic scatter of the linear fitting is higher than 2 dex, and the failure of the thin-disc assumption no longer allows us to measure the dynamical mass for these galaxies. We note that some galaxies have lower dynamical mass than stellar mass, and the differences are not within the $1-\sigma$ error bars, which is likely due to the error bars in the figures being underestimated compared to the true uncertainties for these galaxies. We therefore emphasises again that the interpretations of our results are based on the average of different types of galaxies and not for individual galaxies.

\begin{figure}
    \centering
    \includegraphics[width=\columnwidth]{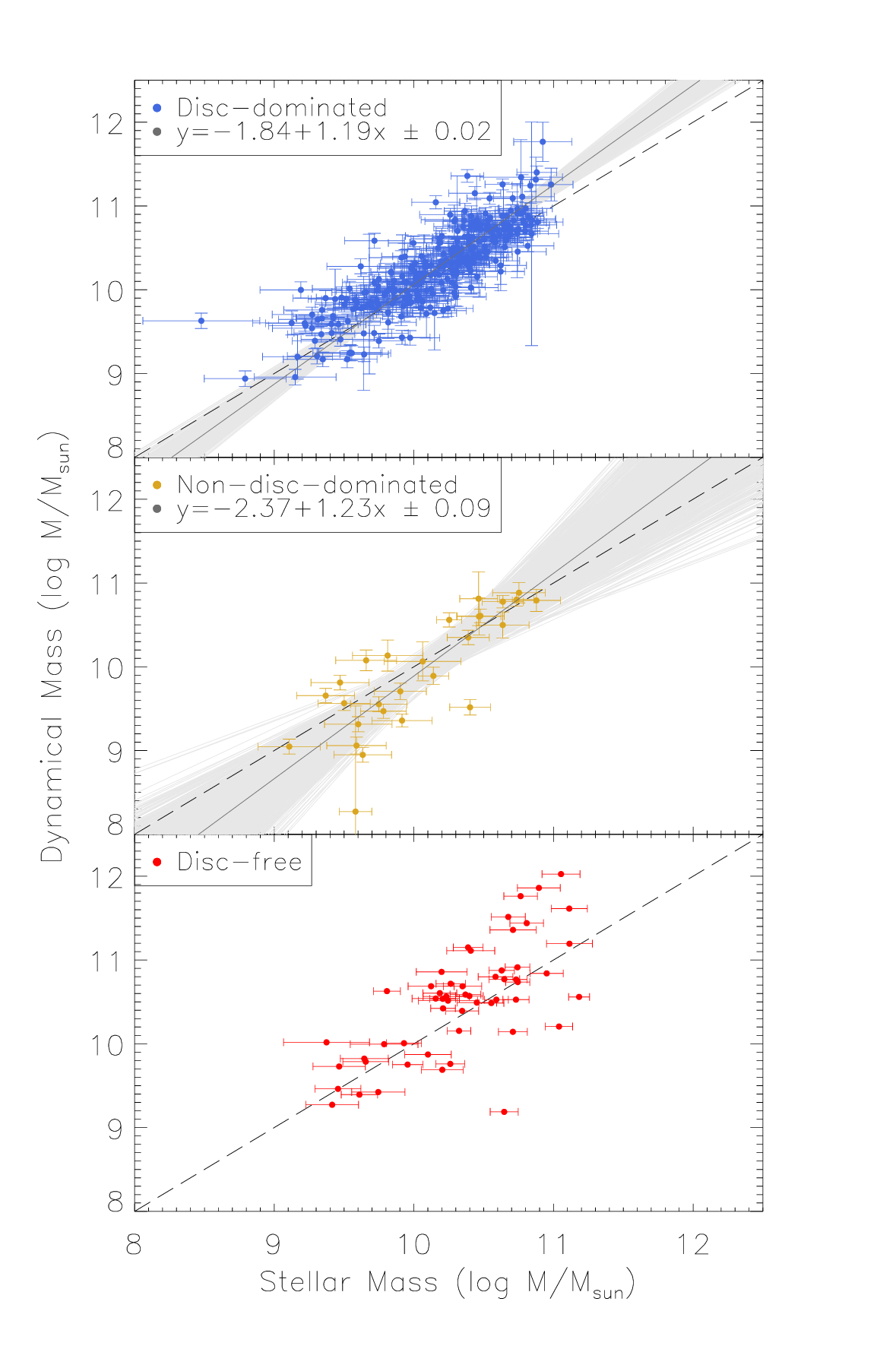}
    \caption{Dynamical mass versus stellar mass (in logarithmic scale) excluding the central area within a radius of $1/4$ side-length of the IFU. From top to bottom: disc-dominated galaxies, non-disc-dominated galaxies and disc-free galaxies. For disc-dominated and non-disc-dominated galaxies, we also fit a linear regression between the logarithmic stellar and dynamical mass as shown in the legends (the dark grey solid line), and light grey lines are the corresponding uncertainty range including the uncertainties of regression coefficients (slope and intercept). The black dashed lines in each panel are the 1-to-1 lines.}
    \label{fig:mass-nb}
\end{figure}

We notice that the disc dynamical mass is higher than the stellar mass on average for our disc-dominated galaxies in the high-mass end in the top panel of Figure~\ref{fig:allmass}.
We first note that the disc dynamical mass density might not trace the real mass density in the inner region of the galaxy because of a possible failure of thin-disc assumption, which might affect the correlation between the disc dynamical mass and the stellar mass for the disc-dominated galaxies.
However, we find that the same correlation between the disc dynamical and stellar masses still exists when we exclude the central area within a radius of $1/4$ of the side-length of the IFU from the mass calculation, as shown in Figure~\ref{fig:mass-nb}. This result indicates the discrepancy of the mass densities in the inner region is negligible for explaining the difference between the dynamical and stellar masses for the disc-dominated galaxies, so we turn to other reasons to explain this difference.

\begin{figure}
     \centering
     \includegraphics[width=\columnwidth]{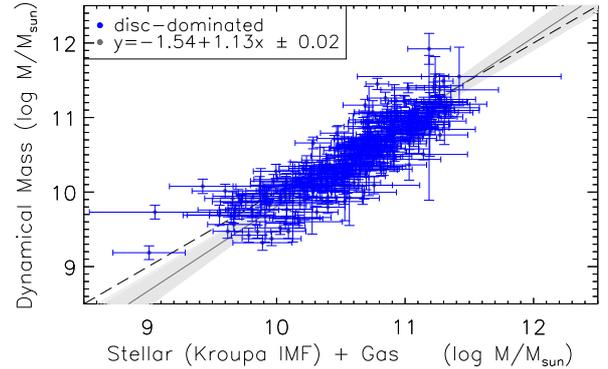}
     \caption{Dynamical mass versus baryonic mass (in logarithmic scale) within MaNGA coverage. The baryonic mass contains stellar mass (Kroupa IMF), atomic and molecular gas mass. We also fit a linear relation for the two masses as shown in the legends (the dark grey solid line), and light grey lines are the corresponding uncertainty range. The black dashed line is the 1-to-1 line.}
     \label{fig:gas}
 \end{figure}

The total baryonic mass contains not only stellar mass but also the mass of other baryonic matter, such as gas and dust, which are not negligible in spiral galaxies. Therefore, we first examine whether the dynamical mass is consistent with the total baryonic mass by including the atomic and molecular gas.
We adopt the gas mass within the coverage of MaNGA estimated from the dust attenuation in the MaNGA PIPE3D value added catalogue~\citep[PIPE3D VAC;][]{2018RMxAA..54..217S}, which are obtained with the \texttt{PIPE3D} pipeline~\citep{2016RMxAA..52...21S,2016RMxAA..52..171S} including the mass of \HI and $\mathrm{H_2}$ estimated with the method described in~\citet{2020MNRAS.492.2651B}.
According to~\citet{2013A&A...557A.131M}, the total atomic and molecular gas mass density is connected to the \HI and $\rm H_2$ density by
\begin{align}
    \Sigma_\mathrm{atom} = 1.4 \Sigma_\mathrm{H\,I}, \\
    \Sigma_\mathrm{mol} = 1.4 \Sigma_\mathrm{H_2}.
\end{align}
Therefore, we adopt the total gas mass by multiplying a factor of $1.4$ to the gas mass obtained from the PIPE3D VAC. The result of this comparison between the total baryonic mass and dynamical mass is shown in Figure~\ref{fig:gas}. 
The dynamical and total baryonic mass are comparable, close to the 1-to-1 line. However, the slope of the correlation between dynamical and total baryonic mass is still larger than 1, which suggests the gas might not be the only reason behind the difference between the dynamical and stellar mass.

Dark matter is present in galaxies in the form of haloes and contributes to the total mass budget, but we do not include a dark matter halo in our model. The dynamical mass we obtained only measures the dynamical mass of the disc within the MaNGA coverage, where the contribution of dark matter is negligible. Figure~\ref{fig:gas} shows that dark matter has a negligible contribution to the dynamical mass of the disc within the MaNGA coverage if we take other baryonic mass into consideration.

Another possible explanation for the discrepancy between dynamical and stellar mass lies with the initial mass function (IMF). The stellar masses we have adopted are measured with the Kroupa IMF for all galaxies, however, the IMFs of galaxies might be mass-dependent such that massive early-type galaxies have an excess of low-mass stars compared to the prediction of the Kroupa IMF, resulting in a bottom-heavy IMF~\citep{2010Natur.468..940V}. Figure~\ref{fig:gas} suggests the possibility of a very slight trend that high-mass disc-dominated galaxies prefer a more bottom heavy IMF which produces a larger number of low-mass stars.

The LSF effect causes a systematic overestimation of velocity dispersion parameters and the resulting mass model. To establish this LSF effect on the disc dynamical mass measurements and the mass relation, we construct mock velocity dispersion profiles by taking average of disc-dominated sample galaxies in three mass bins, and correct the LSF effect for the mock dispersion profiles according to the results from~\citet{2021AJ....161...52L}. We find the dynamical mass for galaxies in the disc-dominated sample is overestimated by about 15 per cent. This overestimation is mass dependent and ranges from 5 per cent for the highest mass galaxies (stellar mass $M_* >10^{11} M_\odot$), to 40 per cent for the lowest mass galaxies (stellar mass $M_* < 10^{9.5} M_\odot$). Considering this mass-dependent systematic error, we estimate that the slope in Figure~\ref{fig:allmass} would be steeper by about 2 percent, within $1-\sigma$ uncertainty due to sample variance and random errors. If we assume the mass discrepancy is caused by an IMF gradient, this systematic error leads to a higher IMF gradient than that is shown in Figure~\ref{fig:gas}.

\subsection{Morphology}
\label{subsec:morph}
As described in Section~\ref{sec:modelfitness}, we classify the sample into three types with two kinematic parameters $\zeta$ and $\eta$: disc-dominated galaxies ($\zeta \leqslant 0.4, \eta \leqslant 1.0$), non-disc-dominated galaxies ($\zeta \leqslant 0.4, \eta > 1.0$) and disc-free galaxies ($\zeta > 0.4$). The disc-free galaxies are the galaxies without a thin disc detection, while the disc-dominated and the non-disc-dominated galaxies describe whether the detected thin disc takes an dominant role or not, respectively.

We divide the sample into spirals, S0s and ellipticals according to the Deep Learning catalogue \citep{2018MNRAS.476.3661D} and the Galaxy Zoo catalogue \citep{2013MNRAS.435.2835W} with the definitions shown in Table~\ref{tab:def}. 
\begin{table*}
  \centering
  \begin{tabular}{l|ll}
  \hline
             & Deep Learning  & Galaxy Zoo     \\ \hline
  Spiral     & TType > 0.0  & t01\_a01\_debiased < 0.8 \& t04\_a08\_debiased $\geqslant$ 0.5 \\
  S0         & TType $\leqslant$ 0.0 \& P\_S0 > 0.5 & t01\_a01\_debiased < 0.8 \& t04\_a08\_debiased < 0.5 \\
  Elliptical & TType $\leqslant$ 0.0 \& P\_S0 $\leqslant$ 0.5 & t01\_a01\_debiased $\geqslant$ 0.8  \\ \hline             \end{tabular}
  \caption{Morphology definition of the two morphology catalogues. 
  For the Deep Learning catalogue: TType > 0 for spiral galaxies; TType $\leqslant$ 0 for S0 and elliptical galaxies; P\_S0 stands for the probability of being a S0 rather than a pure elliptical for galaxies with TType $\leqslant$ 0.
  For the Galaxy Zoo catalogue: t01\_a01\_debiased is the debiased vote fraction for the question if the galaxy is smooth; t04\_a08\_debiased is the debiased vote fraction for the question if the galaxy has spiral arms.}
  \label{tab:def}
\end{table*}

\begin{figure}
    \centering
    \includegraphics[width=\columnwidth]{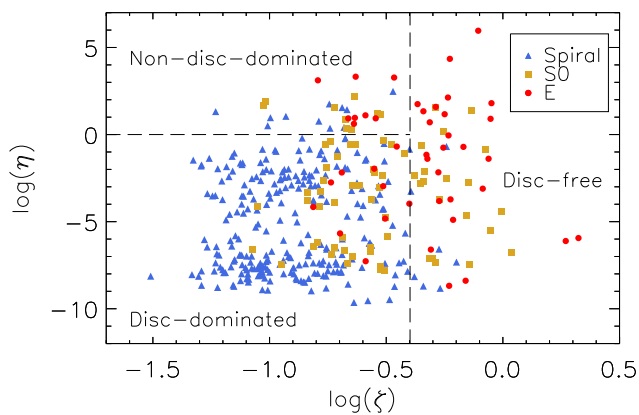}
    \caption{The $\zeta-\eta$ distribution for galaxies with different morphologies. Blue triangles stand for spiral galaxies, yellow squares stand for S0 galaxies and red dots stand for elliptical galaxies. The vertical line ($\zeta = 0.4$) and horizontal ($\eta = 1$) are used to classify our sample into three kinematic types.}
    \label{fig:zeta-eta-morph}
\end{figure}

We first compare our kinematic classification with the morphology classification according to the Deep Learning catalogue in Table~\ref{tab:sta}. We also show the $\zeta-\eta$ distribution for galaxies with different morphologies in Figure~\ref{fig:zeta-eta-morph}.
\begin{table}
\centering
\begin{tabular}{l|lll}
\hline
                & Spiral & S0 & E  \\ \hline
Disc-dominated   & 245    & 40 & 10 \\ 
Non-disc-dominated     & 11     & 9  & 8  \\ 
disc-free    & 7     & 20 & 26 \\ \hline
\end{tabular}
\caption{Morphology statistics for the modelled galaxies. The kinematic and photometric morphology are well correlated. }
\label{tab:sta}
\end{table}

In line with expectation, the kinematic and photometric morphology classification are correlated: the majority of spiral galaxies are dominated by a thin disc, while the model fails on most of the elliptical galaxies. There are also exceptions in Table~\ref{tab:sta} for the following reasons. A few spirals failed to be modelled and were classified as disc-free galaxies, because for these galaxies the parameter $\zeta$ for criteria i) was close to the boundary of 0.4 due to their data quality for $\rm H\alpha$ velocity or stellar velocity dispersion and caused misclassification. Most of the disc-dominated galaxies which were classified as ellipticals in the Deep Learning catalogue are S0s or spirals according to the Galaxy Zoo classification, and after visual inspection, we suspect these are wrongly classified in the Deep Learning catalogue. 

\begin{figure*}
    \centering
    \includegraphics[width=\textwidth]{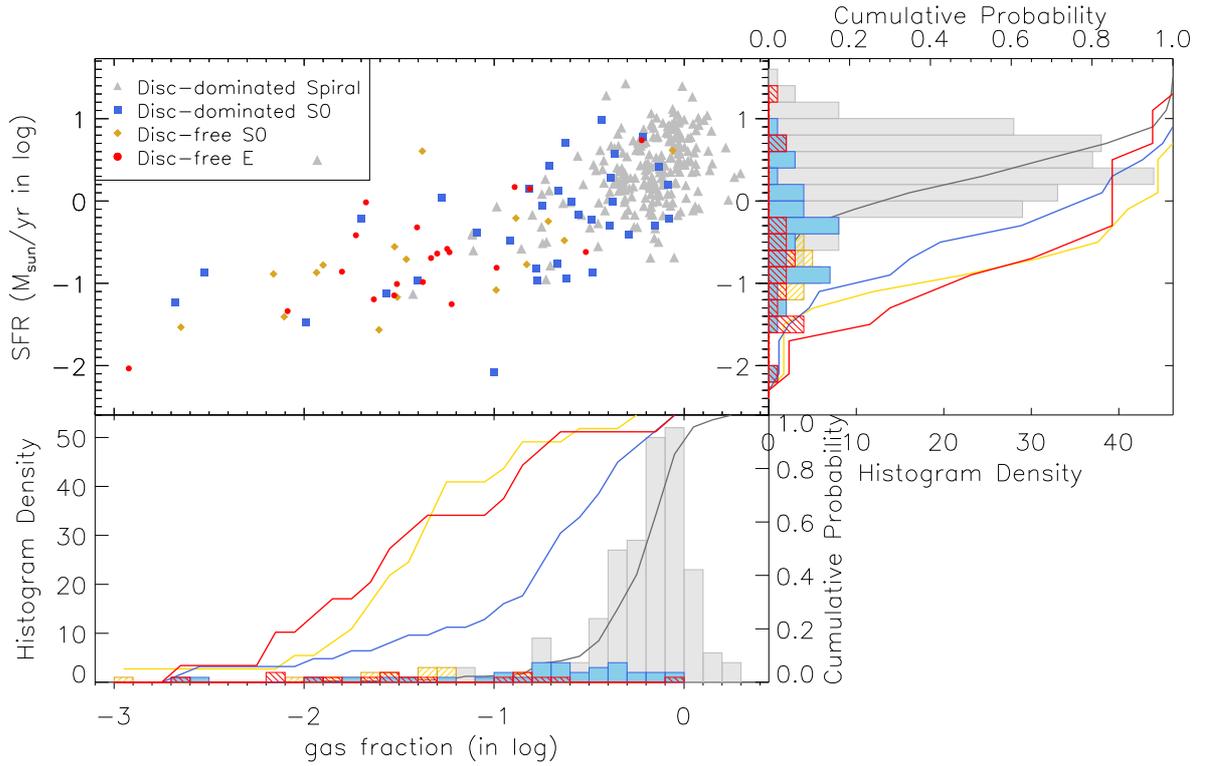}
    \caption{The major panel: the relation between the gas fraction and the SFR. The lower and right panels: the histograms and cumulative possibilities (in solid lines) of the gas fraction and the SFR, respectively. The SFR and molecular gas fraction are positively correlated, which decrease from disc-dominated spirals to disc-dominated S0s to disc-free S0s and disc-free ellipticals.}
    \label{fig:gasfrac-SFR}
\end{figure*}

Table~\ref{tab:sta} shows a clear dichotomy of S0 galaxies. The majority of S0 galaxies are dominated by a thin disc while there are no thin discs detected in some others.
To look for the reason of this dichotomy, we investigate various galaxy properties and notice in two properties major differences between disc-free S0s and disc-dominated S0s. The first property is the integrated star formation rate (SFR) derived from the amount of stellar mass formed in the last 32 Myr in the PIPE3D VAC, and the second property is the gas fraction defined as the ratio between the gas and stellar mass within the MaNGA IFU coverage obtained in Section~\ref{subsec:dm}. We show the histograms and cumulative probabilities of the gas fraction and the SFR, together with their relation, in Figure~\ref{fig:gasfrac-SFR}. 
In general, the SFR and the gas fraction decrease from disc-dominated spirals to disc-dominated S0s to disc-free S0s, while the disc-free ellipticals have a comparable SFR and gas fractions to the disc-free S0s. We find no clear differences in their distributions of other galaxy properties, including their stellar mass and some quantification of galaxy environments as presented in the Galaxy Environment for MaNGA value added catalogue~\citep{2015A&A...578A.110A,2015MNRAS.451..660E}. 

Since molecular gas dominates the gas mass in the inner part of galaxies~\citep{2008AJ....136.2782L,2009AJ....137.4670L}, the gas fraction within the MaNGA coverage ($\sim1.5 R_\mathrm{e}$) in Figure~\ref{fig:gasfrac-SFR} is majorly contributed by the molecular gas. Therefore, the depletion of molecular gas is probably accounting for the quenching process in these S0 galaxies, which is consistent with the conclusion of~\citet{2019ApJ...884L..52Z} that massive disc galaxies are quenched due to the removal of molecular gas content. The removal of molecular gas is possibly a result of consumption due to star formation, AGN or stellar feedback~\citep[e.g.][]{2012ARA&A..50..455F,2014MNRAS.445..581H}. 
Galaxy environments (e.g. ram-pressure stripping, harassment and strangulation) are supposed to play an important role in galaxy quenching~\citep[e.g.][]{1999MNRAS.308..947A,2000Sci...288.1617Q}. We cannot rule out the possibility that molecular gas is removed by processes related to galaxy environments according to our results, for the galaxy environments here are the present environments instead of the past ones. 
The possibility that molecular gas is removed by violent processes (e.g. major merger) is not ruled out as well, though we have a well-defined sample with regular kinematics. Simulations have shown that S0 remnants of major merger still have relaxed structure, leaving almost no discernible morphological traces only 1-2 Gyr after the merger~\citep[][]{2018A&A...617A.113E}.

The kinematic similarity between disc-dominated S0s and spirals is consistent with the theory that disc-dominated S0s formed from spirals with molecular gas removed~\citep{2006A&A...458..101A,2007A&A...470..173B}, and the similarity between disc-dominated S0s and fast-rotating ellipticals also suggests a common origin. However, there are two possible interpretations to the different kinematics between disc-dominated and disc-free S0s. The key question is whether disc-dominated S0s and disc-free S0s originate from different mechanisms, or they are different stages of the same evolutionary path. In the first scenario, it is not implausible to attribute the different kinematics between disc-dominated S0s and disc-free S0s to different mechanisms, for example, disc-free S0s having experienced a more violent quenching process than disc-dominated S0s. The second scenario assumes that disc-dominated S0s and disc-free S0s are different stages of the same mechanism, which suggests the following process: Thin discs can remain stable in S0 galaxies for some time after the molecular gas fraction starts to decrease. Then thin disc structures in S0 galaxies are no longer stable during the exhaustion of molecular gas, leading to the appearance of the disc-free S0s without detectable thin disc structures. Finally, the molecular gas fraction of the disc-free S0s are comparable to those of disc-free ellipticals, marking the end of this quenching process.

\section{Summary}
\label{sec:summary}
In this paper, we have built an analytic model that is fitted to the kinematic maps of stellar and gaseous velocity fields and stellar velocity dispersion field. The model resolves the existence of a thin disc based on the asymmetric drift correction and allows us to measure the disc dynamical mass density. We introduce two parameters $\zeta$ and $\eta$ to describe the fitness of this model and finally classify the sample into disc-dominated, disc-free and non-disc-dominated galaxies. We summarise the results of this work as follows:

\begin{itemize}

\item The $\lambda_{R_\mathrm{e}} - \epsilon$ diagram supports this classification of galaxy morphology. Galaxies dominated by a thin disc are dynamically different from thin-disc-free galaxies, especially fast-rotating ellipticals.

\item The difference between the dynamical and stellar mass of disc-dominated galaxies could be explained in various ways. The explanation of this discrepancy includes taking account of the mass of atomic and molecular gas, as well as varying the IMF. The contribution of dark matter to the dynamical mass of the dominating disc within the MaNGA coverage is negligible.

\item We study the morphology of the sample and find a clear dichotomy in the classification of S0 galaxies. Some S0s are disc-dominated while we fail to detect thin discs in disc-free S0s. We also find the SFR and the gas fraction have a descending trend from disc-dominated spirals to disc-dominated S0s and then to disc-free S0s, and the SFR and the gas fraction of disc-free S0s and disc-free ellipticals are comparable. We propose two scenarios to explain the differences between disc-dominated and disc-free S0s. In the first scenario, disc-free S0s formed in more violent mechanisms than disc-dominated S0s, while in the second scenario, disc-free S0s evolved from disc-dominated S0s as thin-disc structures are not stable after the removal of the molecular gas.
 
\end{itemize}

Our analytical kinematic model provides a simple method to identify thin discs in galaxies. Despite its limitation in decomposing the other components (e.g bulges, bars and thick discs), we obtain robust results related to kinematic structure and mass density for the galaxies in our sample. The application of the model in this paper shows the efficiency of analytic models, which makes them effective in analysing qualitative kinematic data of large integral-field spectroscopic surveys.

\section*{Acknowledgements}
We thank the anonymous referee for their comments to improve this manuscript.

We thank David Law and Kyle Westfall for sharing their analysis of the MaNGA line-spread function in MPL-10 prior to publication. We thank Eric Emsellem for the inspiring discussion about galaxy environments.

MY gratefully acknowledges the financial support from China Scholarship Council (CSC). MAB acknowledges support from NSF/AST-1517006. 

Funding for the Sloan Digital Sky Survey IV has been provided by the Alfred P. Sloan Foundation, the U.S. Department of Energy Office of Science, and the Participating Institutions. SDSS-IV acknowledges
support and resources from the Center for High-Performance Computing at
the University of Utah. The SDSS web site is www.sdss.org.

SDSS-IV is managed by the 
Astrophysical Research Consortium 
for the Participating Institutions 
of the SDSS Collaboration including 
the Brazilian Participation Group, 
the Carnegie Institution for Science, 
Carnegie Mellon University, Center for 
Astrophysics | Harvard \& 
Smithsonian, the Chilean Participation 
Group, the French Participation Group, 
Instituto de Astrof\'isica de 
Canarias, The Johns Hopkins 
University, Kavli Institute for the 
Physics and Mathematics of the 
Universe (IPMU) / University of 
Tokyo, the Korean Participation Group, 
Lawrence Berkeley National Laboratory, 
Leibniz Institut f\"ur Astrophysik 
Potsdam (AIP),  Max-Planck-Institut 
f\"ur Astronomie (MPIA Heidelberg), 
Max-Planck-Institut f\"ur 
Astrophysik (MPA Garching), 
Max-Planck-Institut f\"ur 
Extraterrestrische Physik (MPE), 
National Astronomical Observatories of 
China, New Mexico State University, 
New York University, University of 
Notre Dame, Observat\'ario 
Nacional / MCTI, The Ohio State 
University, Pennsylvania State 
University, Shanghai 
Astronomical Observatory, United 
Kingdom Participation Group, 
Universidad Nacional Aut\'onoma 
de M\'exico, University of Arizona, 
University of Colorado Boulder, 
University of Oxford, University of 
Portsmouth, University of Utah, 
University of Virginia, University 
of Washington, University of 
Wisconsin, Vanderbilt University, 
and Yale University.

%%%%%%%%%%%%%%%%%%%%%%%%%%%%%%%%%%%%%%%%%%%%%%%%%%
\section*{Data availability}
The MaNGA data underlying in this article are publicly available at \url{https://www.sdss.org/dr15/}.

The data generated by our model are available at \url{https://doi.org/10.17630/71cd0f94-5758-4fcc-8e68-fdcd0ff19f7d}.

%%%%%%%%%%%%%%%%%%%% REFERENCES %%%%%%%%%%%%%%%%%%

\bibliographystyle{mnras}
\bibliography{morph}

%%%%%%%%%%%%%%%%%%%%%%%%%%%%%%%%%%%%%%%%%%%%%%%%%%
%%%%%%%%%%%%%%%%% APPENDICES %%%%%%%%%%%%%%%%%%%%%
%\appendix

% Don't change these lines
\bsp	% typesetting comment
\label{lastpage}
\end{document}